\documentclass[a4paper,fleqn,usenatbib]{mnras}

\usepackage{newtxtext,newtxmath}
\usepackage[T1]{fontenc}
\usepackage{ae,aecompl}

\usepackage{graphicx}	% Including figure files
\usepackage{amsmath}	% Advanced maths commands
\usepackage{amssymb}	% Extra maths symbols

\newcommand{\mpl}{M_{\rm p}}
\newcommand{\mstar}{M_{\ast}}
\newcommand{\rp}{R_{\rm p}}

\newcommand{\rhosurf}{\Sigma}

\newcommand{\hp}{h_{\rm p}}

\newcommand{\sigmaunp}{\rhosurf_{\rm un,p}}

\newcommand{\omegakp}{\Omega_{\rm K,p}}

\defcitealias{Kanagawa_Tanaka_Szuszkiewicz2018}{KTS18}
\defcitealias{Scardoni19}{S20}
\defcitealias{Durmann_Kley2015}{DK15}

\newcommand{\REVI}[1]{#1}
\newcommand{\REVII}[1]{#1}
\title[Comments on ``Type II migration strikes back'']{Comments on ``Type II migration strikes back -- An old paradigm for planet migration in discs'' by Scardoni et al.}

\author[K. D. Kanagawa]{
Kazuhiro D. Kanagawa$^{1}$\thanks{E-mail:kazuhiro.kanagawa@utap.phys.s.u-tokyo.ac.jp},
and Hidekazu Tanaka$^{2}$
\\
$^{1}$Research Center for the Early Universe, Graduate School of Science, The University of Tokyo, Hongo, Bunkyo-ku, Tokyo 113-0033, Japan \\
$^{2}$Astronomical Institute, Tohoku University, Sendai, Miyagi 980-8578, Japan
}

\date{Accepted 2020 April 8. Received 2020 April 8; in original form 2020 January 13}

\pubyear{2020}

\begin{document}
\label{firstpage}
\pagerange{\pageref{firstpage}--\pageref{lastpage}}
\maketitle

\begin{abstract}
In the conventional view of type~II migration, a giant planet migrates inward in the viscous velocity of the accretion disc in the so-call disc-dominate case.
Recent hydrodynamic simulations, however, showed that planets migrate with velocities much faster than the viscous one in massive discs.
Such fast migration cannot be explained by the conventional picture.
\cite{Scardoni19} has recently argued this new picture.
By carrying out similar hydrodynamic simulations, they found that the migration velocity slows down with time and eventually reaches the prediction by the conventional theory.
They interpreted the fast migration as an initial transient one and concluded that the conventional type II migration is realized after the transient phase.
We show that the migration velocities obtained by \cite{Scardoni19} are consistent with the previous simulations even in the transient phase that they proposed.
We also find that the transient fast migration proposed by \cite{Scardoni19} is well described by a new model of \cite{Kanagawa_Tanaka_Szuszkiewicz2018}.
The new model can appropriately describe significant inward migration during the initial transient phase that \cite{Scardoni19} termed.
Hence, we conclude that the time-variation of the transient migration velocity is due to the changes of the orbital radius of the planet and its background surface density during the migration.
\end{abstract}


%%%%%%%%%%%%%%%%%%%%%%%%%%%%%%%%%%%%%%%%%%%%%%%%%%

%%%%%%%%%%%%%%%%% BODY OF PAPER %%%%%%%%%%%%%%%%%%
\section{Introduction} \label{sec:intro}
A planet formed within a protoplanetary disc interacts with the surrounding gas and migrates as a consequence of the disc--planet interaction \citep[e.g.,][]{Goldreich_Tremaine1980}.
When it is massive enough, the planet forms a density gap along with its orbit and migrates together with the gap, which is the so-called Type~II migration \citep[e.g.,][]{Lin_Papaloizou1979,Armitage2007}.
In the conventional view of the type~II migration, in the massive disc, the planet migration is locked into viscous evolution of the disc and migrates inward in the same velocity of the surrounding gas (i.e., the so-called disc-dominate case).
However, recent hydrodynamic simulations found that the inward migration velocity of the giant planet can be faster than the velocity of the gas viscous drift velocity \citep[e.g.,][]{Duffell_Haiman_MacFadyen_DOrazio_Farris2014,Durmann_Kley2015,Kanagawa_Tanaka_Szuszkiewicz2018,Robert_Crida_Lega_Meheut_Morbidelli2018}, which is the evidence of that the mechanism of the migration is different from that supposed by the conventional type II migration.
In this context, \cite{Kanagawa_Tanaka_Szuszkiewicz2018} (hereafter \citetalias{Kanagawa_Tanaka_Szuszkiewicz2018}) have indicated that the torque on the planet can be described by a torque formula similar to the type~I by using the surface density at the bottom of the gap and the giant planet migrates according to the torque.
\REVI{Moreover, \cite{Robert_Crida_Lega_Meheut_Morbidelli2018} have shown that the giant planet can migrate inward together with the gap even in a disc with a zero accretion rate.
The above results indicate that the migration of the giant planet is driven by the torque exerted from the surrounding gas, rather than the viscosity.
}

Recently \cite{Scardoni19} (hereafter \citetalias{Scardoni19}) have argued the migration of a giant planet in the massive disc.
By carrying out hydrodynamic simulations with a Jupiter-mass planet, \citetalias{Scardoni19} have found that the velocity of the migration gradually decreases and eventually reaches the velocity predicted by the conventional type II migration, though it is much faster than the velocity of the gas viscous drift in an initial phase.
\REVII{\citetalias{Scardoni19} interpreted the observed fast migration as an initial transient one and concluded that the conventional type II migration will be realized after the initial transient.}

However, the time-variation of the migration velocity can be due to the change of the parameters in the migration formula, for instance, the planetary orbital radius $\rp$ and the unperturbed surface density at $\rp$, $\sigmaunp$.
\cite{Duffell2019b} investigated gas accretion onto the secondary of a binary system and its orbital migration for a wide range of the binary mass ratio from $0.01$ to unity.
To do so, they employ a technique by which the parameter space is scanned continuously, by slowly changing the mass ratio with time.
This technique enables them to thoroughly explore the parameter space in only a few numerical runs.
Here we clarify whether the transient migration can be described by a new model of \citetalias{Kanagawa_Tanaka_Szuszkiewicz2018}, due to the change of the parameters during the migration.

\section{Comparison of migration velocities} \label{sec:results}
From the hydrodynamic simulations of \cite{Durmann_Kley2015} (hereafter \citetalias{Durmann_Kley2015}) and \citetalias{Kanagawa_Tanaka_Szuszkiewicz2018}, the migration velocity normalized by the viscous velocity of the gas is obtained as a function of $\sigmaunp \rp^2/\mpl$, where $\rp$ and $\mpl$ are the orbital radius and the mass of the planet respectively, and $\sigmaunp$ denotes the unperturbed surface density at $R=\rp$.
\citetalias{Kanagawa_Tanaka_Szuszkiewicz2018} gives the empirical formula as
\begin{align}
\frac{u_{\rm p}}{u_{\rm vis}} &= c \frac{(\hp/\rp)}{0.03}\frac{\sigmaunp \rp^2}{\mpl},
\label{eq:k18_formula}
\end{align}
where $c$ is the fitting parameter and \REVII{$c$ takes $1$ -- $3$}, $\hp$ is the disc scale height at $R=\rp$, and the radial velocity of the gas viscous drift is given by
\begin{align}
u_{\rm vis} = - \frac{3}{2} \frac{\nu_{\rm p}}{\rp} 
\end{align}
where $\nu_{\rm p}$ denotes the kinetic viscosity at the planet orbit $\rp$.
For a comparison with the results of \citetalias{Scardoni19}, we introduce
\begin{align}
B=4\pi \frac{\sigmaunp \rp^2}{\mpl}.
\label{eq:b}
\end{align}

\begin{figure*}
\includegraphics[width=\textwidth]{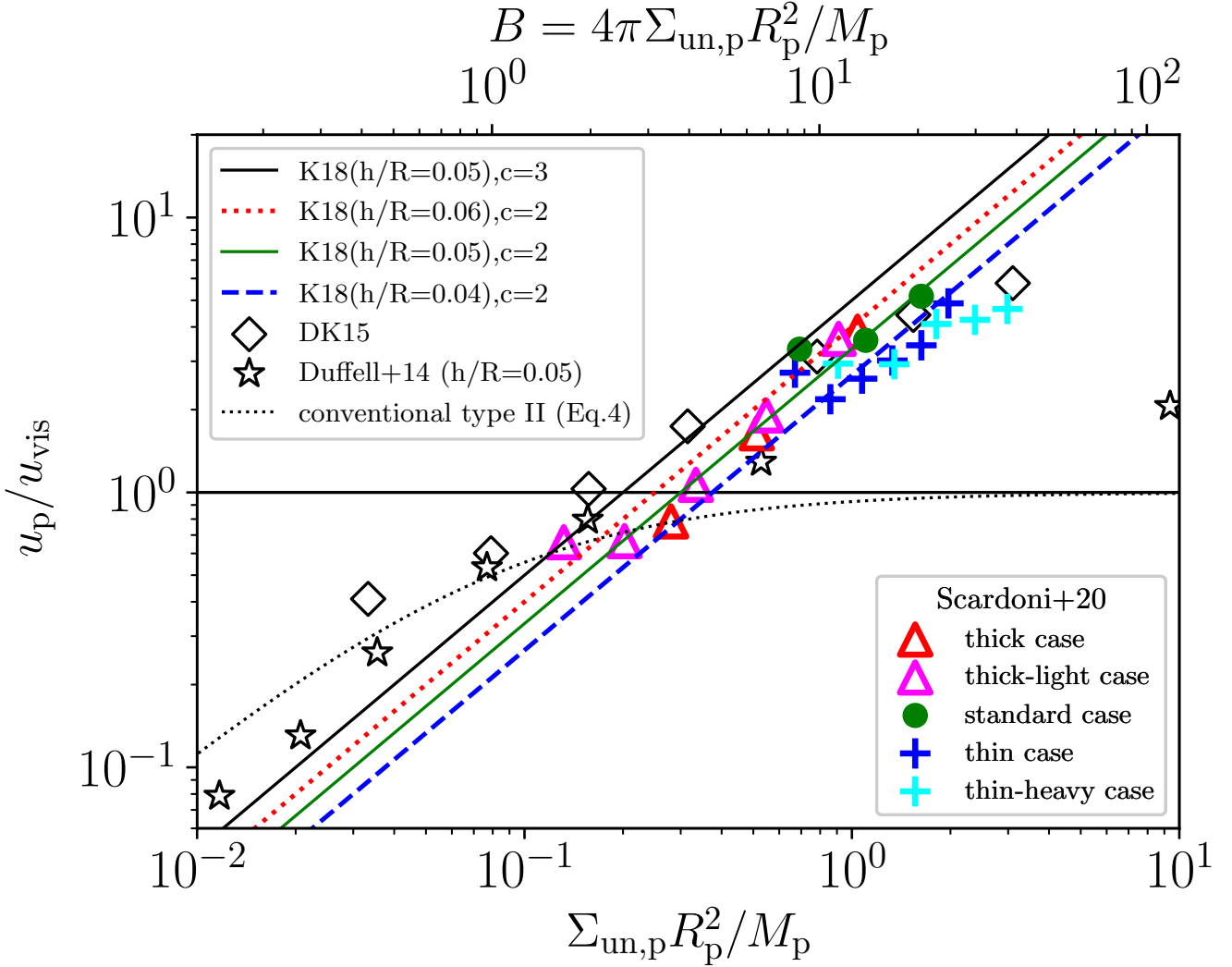}
\caption{
Ratio of the inward migration velocity $u_{\rm p}$ to the viscous velocity $u_{\rm vis}$ as a function of the value $\sigmaunp \rp^2/\mpl$ (and the upper x-axis is $B=4\pi \sigmaunp \rp^2/\mpl$).
The crosses, triangles, and circles indicate the data extracted from \protect \citetalias{Scardoni19}.
The time progresses as $\sigmaunp \rp^2/\mpl$ decreases.
That is, from the right, the data at $t=500$~orbit, $t=1000$~orbit, and $t=1500$~orbit.
For the thin, thin-heavy and thick-light cases, the data at the later phase are plotted in the further right.
The diamonds and stars indicate the results given by \protect \citetalias{Durmann_Kley2015} and \protect \cite{Duffell_Haiman_MacFadyen_DOrazio_Farris2014}, respectively.
The dotted and thin solid, dashed lines are the empirical formula of Equation~(\ref{eq:k18_formula}) with $c=2$ for $\hp/\rp=0.06$, $0.05$ and, $0.04$ from the top, respectively.
The thick solid line indicates the Equation~(\ref{eq:k18_formula}) with $c=3$ for $\hp/\rp=0.05$.
The thin dotted line indicates the prediction of the conventional type~II migration of Equation~(\ref{eq:uii}).
}
\label{fig:comp_migvel}
\end{figure*}
\citetalias{Scardoni19} showed time variations of the ratio of the inward migration velocity of the planet to the migration velocity predicted by the conventional type II migration ($u_{\rm II}$) and the value of $B$:
$u_{\rm II}$ is given by
\begin{align}
u_{\rm II} &= u_{\rm vis} \frac{B}{B+1}.
\label{eq:uii}
\end{align}
\citetalias{Scardoni19} simulated the migration of the Jupiter-mass planet with three different values of $h/R$: $h/R=0.05$ (standard case), $h/R=0.04$ (thin case) and  $h/R=0.06$ (thick case).
In addition, they carried out the simulation with a heavier disc (the initial surface density is $\simeq 1.56$ times larger than that in the standard case) with $h/R=0.04$ (thin-heavy case) and the simulation with a lighter disc (the initial surface density is $\simeq 0.69$ times smaller) with $h/R=0.06$ (thick-light case).
We extracted the $u_{\rm p}/u_{\rm II}$ and $B$ for each case from Figures~2, 3 and 5 of \citetalias{Scardoni19} and plot the values at $t=500$~orbit, $1000$~orbit, and $1500$~orbit in Figure~\ref{fig:comp_migvel} (Note that we converted $u_{\rm II}$ to $u_{\rm vis}$ by using Equation~(\ref{eq:uii})).
In addition, we plot the data at $t=2000$~orbit and $2500$~orbit for the thin-heavy and thick-light case, and for thin case, the data at $t=2000$~orbit, $2500$~orbit and $3000$~orbit are plotted.
\REVII{
In Figure~\ref{fig:comp_migvel}, we also plot the results given by \cite{Duffell_Haiman_MacFadyen_DOrazio_Farris2014}\footnote{Note that when $\sigmaunp \rp^2/\mpl \gg 1$, the migration velocities given by \cite{Duffell_Haiman_MacFadyen_DOrazio_Farris2014} are much slower than those given by other works. This discrepancy could be due to that \cite{Duffell_Haiman_MacFadyen_DOrazio_Farris2014} adopted the different way to measure the migration velocity.} and \citetalias{Durmann_Kley2015}, and the lines predicted by the model of \citetalias{Kanagawa_Tanaka_Szuszkiewicz2018} with $c=2$ and $c=3$.
}

As \citetalias{Scardoni19} showed, the migration velocity of the planet slows down with time.
However, the time-variation of the migration velocity varies with time along with the line predicted by \REVII{the model of \citetalias{Kanagawa_Tanaka_Szuszkiewicz2018} with $c=2$.
Their results are also consistent with those given by \citetalias{Durmann_Kley2015}.}
For instance, in the standard case, the migration velocity is quite similar to that given by \citetalias{Durmann_Kley2015} at the similar $\sigmaunp \rp^2/\mpl$.
\REVI{
Note that we also comfirmed that the evolution of the semi-major axis is similar to the sdandard case of \citetalias{Scardoni19} and the model of \citetalias{Durmann_Kley2015} with $\dot{m}=10^{-7} M_{\odot}/\rm{yr}$ (corresponds $\sigma_0=2.7\times 10^{-3}$, which is equivalent to that in the standard case) shown in Figure~6 of \citetalias{Durmann_Kley2015}.
For instance, at $t=500$~orbit, $\rp\simeq 0.7$ and $\rp\simeq 0.55$ around $t=1000$~orbit.
}

As time progresses, the migration velocity slows down but this velocity approaches to the line predicted by \citetalias{Kanagawa_Tanaka_Szuszkiewicz2018} (Equation~\ref{eq:k18_formula}) \REVII{with $c=2$}, rather than the viscous velocity and the prediction of the conventional type~II migration (Equation~\ref{eq:uii}).
In other cases, the migration velocity also varies along with the line of Equation~(\ref{eq:k18_formula}) with time as in the standard case.
Only at the last phases of the thick and thick-light cases, the migration velocity given by \citetalias{Scardoni19} reaches the velocity predicted by the conventional type~II migration when $\sigmaunp \rp^2/\mpl = 0.1$ -- $0.2$.
This is also consistent with the results given by \citetalias{Durmann_Kley2015} and \citetalias{Kanagawa_Tanaka_Szuszkiewicz2018}.
Those previous studies found that the migration velocity is close to the viscous velocity when  $\sigmaunp \rp^2/\mpl = 0.1$ -- $0.2$.
\REVI{Note that the migration significantly speeds up around the last phase in the simulations done by \citetalias{Scardoni19}.
This speed-up could be caused by the effect of the inner boundary, because the gap shape (especially inner structure) at the later phase is clearly affected by the inner boundary as can be seen in Figure~7 of \citetalias{Scardoni19}.}

\REVII{
To fit the data when the disc is massive, namely $\sigmaunp \rp^2/\mpl \gtrsim 1$, we found that in Equation~(\ref{eq:k18_formula}), $c=2$ is better.
On the other hand, for the data in the case of $\sigmaunp \rp^2/\mpl < 1$, the choice of $c=3$ looks better.
This might indicate that the coefficient may be different in the case of $\sigmaunp \rp^2/\mpl > 1$ and the case of $\sigmaunp \rp^2/\mpl < 1$, though this dependence is not related to the theory of the conventional type II migration. }
Also we should note that around $\sigmaunp \rp^2/\mpl \simeq 3$, the velocities given by the both simulations of \citetalias{Durmann_Kley2015} and \citetalias{Scardoni19} are slightly slower than that expected by Equation~(\ref{eq:k18_formula}).
Although the reason of this deviation is not clear, there may be an upper limit of $u_{\rm p}/u_{\rm vis}$ (but this upper limit is larger than that \cite{Duffell_Haiman_MacFadyen_DOrazio_Farris2014} found).

Note that \citetalias{Scardoni19} compared the torque exerted on the gas crossing the gap and the torque exerted on the inner and outer discs and found that the former torque is negligible to the latter one.
However, it does not mean that the torque is not exerted at the bottom of the gap.
\REVII{
In the model of \citetalias{Kanagawa_Tanaka_Szuszkiewicz2018}, the most of the torque is assumed to be exerted from the bottom of the gap, which is related to the Lindblad torque and not related to the corotation torque and torque exerted from the gas crossing the gap.
The bottom of the gap does not mean only the co-orbital region of the planet, it is wider \footnote{\citetalias{Kanagawa_Tanaka_Szuszkiewicz2018} defined the bottom of the gap as the region from $R = \rp - \delta$ to $R = \rp + \delta$ with $\delta = 2\max(R_H, \hp)$, excised from $\psi = \psi_{\rm p} - \delta /\rp$ to $\psi = \psi_{\rm p} + \delta / \rp$ as followed by  \cite{Fung_Shi_Chiang2014}), where $R_H$ is the Hill radius and $\psi_{\rm p}$ indicates the azimuthal angle of the planet.}.
As can be seen in Figure~2 of \citetalias{Kanagawa_Tanaka_Szuszkiewicz2018}, the above assumption that the most of the torque is exerted from the bottom of the gap, agrees with the results of hydrodynamic simulations.
\citetalias{Durmann_Kley2015} also obtained the similar results (e.g., Figures~8 and 9 of that paper).
}

\REVI{
\cite{Robert_Crida_Lega_Meheut_Morbidelli2018} showed that even in the disc with the zero accretion rate (case B1), the giant planet migrates inward in the similar velocity to that in the disc with a finite accretion rate (case A1).
\citetalias{Scardoni19} discussed this result in Section 6.3.
Their explanation is that due to the perturbations caused by the planet, the surface density profile no longer satisfies the steady-state accretion disc and these perturbations will quickly modify the surface density profile towards the initial condition of the case of A1.
Because the surface density profile becomes similar, the migration velocity in the initial transient phase is similar.
However, as \cite{Duffell_Haiman_MacFadyen_DOrazio_Farris2014} and \citetalias{Durmann_Kley2015} pointed out, the gap formation does not change the gas accretion rate of the disc (for instance, see Figure~4 of \citetalias{Durmann_Kley2015}).
In the disc with a zero accretion rate, we can expect that the gap formation does not generate a significant mass flux comparable to that in A1, considering the above results.
Hence, the explanation of \citetalias{Scardoni19} seems to be inconsistent with the results of the previous works.
}

\REVI{
In Section 6.3, \citetalias{Scardoni19} also argued that the initial transient cannot be maintained since the gas cannot move faster than the viscous drift velocity and hence the migration velocity slows down because of the depletion of the gas density outside the planet.
This effect was also observed by \citetalias{Durmann_Kley2015} (see Section 4.2 of \citetalias{Durmann_Kley2015}).
However, \citetalias{Durmann_Kley2015} found that the most of the torque is exerted within a $5\hp$ wide region inside and outside the planet, and the surface density distribution of this region does not change after several hundred orbits.
Consequently, the torque also does not change in time.
One can confirm that in Figure~7 of \citetalias{Scardoni19}, the surface density distribution does not change in a $5\hp$ region inside and outside the planet, during the migration.
The above fact is also consistent with the results of \citetalias{Kanagawa_Tanaka_Szuszkiewicz2018} and the assumption of their model mentioned above.
The structure in the vicinity of the planet can reach the quasi-steady state within a timescale of local viscous diffusion, rather than the viscous evolution timescale of entire disc.
Indeed, as \citetalias{Durmann_Kley2015} showed, the gap shape (e.g., Figure 7) and the normalized torque\footnote{To consider a change of a backgraound, the torque should be normalized by $\Gamma_0=(\mpl/\mstar)^2(\hp/\rp)^{-2} \sigmaunp \rp^4 \omegakp^2$} (e.g., Figure~13) hardly depend on time after several hundred orbits.
Moreover, \cite{Kanagawa2017b} showed that the gap width and depth become stationary within the viscous diffusion time across the width. 
In this sense, the model of the previous studies reaches quasi-steady state.
}

\section{Conclusion} \label{sec:conclusion}
We showed that the results of simulations carried out by \citetalias{Scardoni19} are consistent with those given by the previous studies (e.g., \citetalias{Durmann_Kley2015} and \citetalias{Kanagawa_Tanaka_Szuszkiewicz2018}).
The time variation of the fast migration in the transient phase proposed by \cite{Scardoni19} is due to the decrease in the ratio $\sigmaunp \rp^2/\mpl$ (or $B$) during the migration.
The agreement between the planetary migration velocity and the velocity of gas viscous drift is realized when $\sigmaunp \rp^2/\mpl = 0.1$ -- $0.2$, as \citetalias{Durmann_Kley2015} and \citetalias{Kanagawa_Tanaka_Szuszkiewicz2018} already showed.

\REVII{
We also found that the transient fast migration proposed by \citetalias{Scardoni19} is well described by a new model of \citetalias{Kanagawa_Tanaka_Szuszkiewicz2018}, as a function of $\sigmaunp \rp^2/\mpl$.
Giant planets experience significant inward migration during the initial transient phase that \citetalias{Scardoni19} termed.
The new model of \citetalias{Kanagawa_Tanaka_Szuszkiewicz2018} can be applied to the evolution in this phase.
}

\REVI{
The results of \citetalias{Durmann_Kley2015} and \citetalias{Kanagawa_Tanaka_Szuszkiewicz2018} support that the migration of a giant planet is driven by the torque exerted from the surrounding gas, rather than the accretion flow of the disc gas.
The gap shape and the torque can reach quasi-stady states in the local viscous timescale, which is much shorter than the viscous evolution time of the entire disc.
}
Moreover, an initial distribution of the surface density of hydrodynamic simulations is different in each paper.
\citetalias{Durmann_Kley2015} adopted the simple power-law distribution, \citetalias{Kanagawa_Tanaka_Szuszkiewicz2018} used the simple power-law distribution and the power-law distribution with the empirical gap shape.
\citetalias{Scardoni19} used a well-relaxed initial surface density distribution as described in Section~4.1 of that paper.
Nevertheless, as can be seen in Figure~\ref{fig:comp_migvel}, the migration velocity converges to the line predicted by \citetalias{Kanagawa_Tanaka_Szuszkiewicz2018} (Equation~\ref{eq:k18_formula}), regardless of the initial condition.
This convergence also indicates that the planetary migration is in quasi-steady state even in the initial transient phase that \citetalias{Scardoni19} proposed.

\section*{Acknowledgments}
We would like to thank the anonymous referees for useful comments.
K.D.K was supported by JSPS Core-to-Core Program ``International Network of Planetary Sciences'' and JSPS KAKENHI grant 19K14779.
H.T. was supported by JSPS KAKENHI grant 18H05438, and 17H01103.
%%%%%%%%%%%%%%%%%%%%%%%%%%%%%%%%%%%%%%%%%%%%%%%%%%



%\bibliographystyle{mnras}
%\bibliography{reference} % if your bibtex file is called example.bib

% Don't change these lines
\bsp	% typesetting comment
\label{lastpage}
\end{document}